\def \SAIT #1 #2 {{\em Mem.\ Soc.\ Astron.\ It.\/} {\bf #1}, #2}
\def \MESS #1 #2 {{\em The Messenger\/} {\bf #1}, #2}
\def \ASTRNACH #1 #2 {{\em Astron. Nach.\/} {\bf #1}, #2}
\def \AAP #1 #2 {{\em Astron. Astrophys.\/} {\bf #1}, #2}
\def \AAL #1 #2 {{\em Astron. Astrophys. Lett.\/} {\bf #1}, L#2}
\def \AAR #1 #2 {{\em Astron. Astrophys. Rev.\/} {\bf #1}, #2}
\def \AAS #1 #2 {{\em Astron. Astrophys. Suppl. Ser.\/} {\bf #1}, #2}
\def \AJ #1 #2 {{\em Astron. J.\/} {\bf #1}, #2}
\def \ANNREV #1 #2 {{\em Ann. Rev. Astron. Astrophys.\/} {\bf #1}, #2}
\def \APJ #1 #2 {{\em Astrophys. J.\/} {\bf #1}, #2}
\def \APJL #1 #2 {{\em Astrophys. J. Lett.\/} {\bf #1}, L#2}
\def \APJS #1 #2 {{\em Astrophys. J. Suppl.\/} {\bf #1}, #2}
\def \APSS #1 #2 {{\em Astrophys. Space Sci.\/} {\bf #1}, #2}
\def \ASR #1 #2 {{\em Adv. Space Res.\/} {\bf #1}, #2}
\def \BAIC #1 #2 {{\em Bull. Astron. Inst. Czechosl.\/} {\bf #1}, #2}
\def \JSQRT #1 #2 {{\em J. Quant. Spectrosc. Radiat. Transfer\/} {\bf #1}, #2}
\def \MN #1 #2 {{\em Mon. Not. R. Astr. Soc.\/} {\bf #1}, #2}
\def \MEM #1 #2 {{\em Mem. R. Astr. Soc.\/} {\bf #1}, #2}
\def \PLR #1 #2 {{\em Phys. Lett. Rev.\/} {\bf #1}, #2}
\def \PR #1 #2 {{\em Phys. Rev.\/} {\bf #1}, #2}
\def \PASJ #1 #2 {{\em Publ. Astron. Soc. Japan\/} {\bf #1}, #2}
\def \PASP #1 #2 {{\em Publ. Astr. Soc. Pacific\/} {\bf #1}, #2}
\def \NAT #1 #2 {{\em Nature\/} {\bf #1}, #2}

\documentstyle[twoside,psfig]{memsait}

\begin{opening}
\title{The Nuclear Level Density and the Determination of Thermonuclear Rates
     for Astrophysics}
\author{T. Rauscher$^1$, F.-K. Thielemann$^1$, K.-L. Kratz$^2$}
\institute{$^1$Institut f\"ur theoretische Physik, Universit\"at Basel, Basel,
Switzerland\\
$^2$Institut f.\ Kernchemie, Universit\"at Mainz, Mainz, Germany}
\date{} 
\end{opening}

\begin{document}

\def\beginrefer{\section*{References}%
\begin{quotation}\mbox{}\par}
\def\refer#1\par{{\setlength{\parindent}{-\leftmargin}\indent#1\par}}
\def\endrefer{\end{quotation}}

\oddpagefooter{}{}{} 
\evenpagefooter{}{}{} 
\ 
\bigskip

\begin{abstract}
The prediction of cross sections for nuclei far off stability is crucial
in the field of nuclear astrophysics. In recent calculations the nuclear
level density -- as an important ingredient to the statistical model
(Hauser-Feshbach) --
has shown the highest uncertainties.
We present a global parametrization of nuclear level densities based on the
back-shifted Fermi-Gas formalism. Employment of an energy-dependent level
density parameter $a$ and microscopic corrections from a recent FRDM mass
formula by M\"oller et al.\
leads to a highly improved fit of level densities at the
neutron-separation energy in the mass range $20\le A \le 245$. The
importance of using proper microscopic corrections from mass formulae is
emphasized. The resulting level description is well suited for
astrophysical applications.

The level density can also provide clues to the applicability of the
statistical model which is only correct for a high density of excited states. 
Using the above description one can derive a ``map''
for the applicability of the model for reactions of stable and unstable
nuclei with neutral and charged particles.
\end{abstract}

\section{Introduction}

Explosive nuclear burning in astrophysical environments produces unstable 
nuclei, which again can be targets for subsequent reactions. In addition,
it involves a
very large number of stable nuclei, which are not fully explored
by experiments. Thus, it is  necessary to be able to predict reaction cross 
sections and thermonuclear rates with the aid of theoretical models.
Explosive burning in supernovae involves in general intermediate mass
and heavy nuclei. Due to a large nucleon number they have intrinsically
a high density of excited states. A high level density in the 
compound nucleus at the appropriate excitation energy allows to
make use of the statistical model approach for compound nuclear 
reactions (e.g. Hauser \& Feshbach 1952, Mahaux \& Weidenm\"uller 1979), 
which averages over resonances.


As the capture of an alpha particle leads usually to larger Q-values than
neutron or proton captures, the compound nucleus is created at a higher
excitation energy and especially in the case of alpha-captures it
is often even possible to apply the Hauser-Feshbach formalism for light nuclei.
Another advantage of alpha-captures is that the capture Q-values vary
little with the N/Z-ratio of a nucleus, for nuclei with Z$\le$50. For Z$>$50,
entering the regime of natural alpha-decay, very small alpha-capture Q-values
can be encountered for proton-rich nuclei. Such nuclei on the other hand do
not play a significant role in astrophysical environments, maybe with 
exception of the p-process.
This means that in the case of alpha-captures the requirement of large level 
densities at the bombarding energy is equally well fulfilled
at stability as for unstable nuclei. 
Opposite to the behavior for alpha-induced reactions,
the reaction Q-values for proton or neutron captures vary strongly
with the N/Z-ratio, leading eventually to vanishing Q-values at
the proton or neutron drip line. For small Q-values the compound
nucleus is created at low excitation energies and also for intermediate
nuclei the level density can be quite small. Therefore, it is not
advisable to apply the statistical model approach close to the 
proton or neutron drip lines for intermediate nuclei.
For neutron captures close to the neutron drip line in r-process
applications it might be still permissable for heavy and often
deformed nuclei, which have a high level density already at
very low excitation energies. 

In astrophysical applications usually different points are emphasized than 
for investigations in pure nuclear physics. Many of
the latter in this well established field were focused on specific
reactions, where all or most "ingredients", like optical potentials for
particle and alpha transmission coefficients, level densities, resonance
energies and widths of giant resonances to be implementated in predicting
E1 and M1 gamma-transitions, were deduced from experiments. This of course,
as long as the statistical model prerequisits are met, will produce highly
accurate cross sections.
For the majority of nuclei in astrophysical applications such information
is not available. The real challenge is thus not the well established
statistical model, but rather to provide all these necessary ingredients
in as reliable a way as possible, also for nuclei where none of such 
informations are available. In addition, these approaches should be on a
similar level as e.g. mass models, where the investigation of hundreds
or thousands of nuclei is possible with manageable computational effort,
which is not always the case for fully microscopic calculations.

The major uncertainty
in all existing calculations stems from the prediction of nuclear level
densities (Truran et al.\ 1966; Holmes et al.\ 1976; Cowan, Thielemann \& 
Truran 1991, and extended references therein), which in earlier calculations 
gave uncertainties even beyond
a factor of 10 at the neutron separation energy (Gilbert \& Cameron 1965),
about a factor of 8 (Woosley et al. 1978), and a factor of 5 even in
the most recent calculations (e.g. Thielemann, Arnould \& Truran 1987; see 
also Fig.\ 3.16 in Cowan, Thielemann \& Truran 1991). 

\section
{Thermonuclear Rates from Statistical Model Calculations} 


A high level density in the compound nucleus permits to use averaged
transmission coefficients $T$, which do not reflect a resonance behavior,
but rather describe absorption via an imaginary part in the (optical)
nucleon-nucleus potential (for details see Mahaux \&
Weidenm\"uller 1979). This leads to the well known expression
\begin{eqnarray}
\sigma^{\mu \nu}_{i} (j,o;E_{ij})& = &
{{\pi \hbar^2 /(2 \mu_{ij} E_{ij})} \over 
{(2J^\mu_i+1)(2J_j+1)}} \nonumber \\
 & & \times \sum_{J,\pi} (2J+1){{T^\mu_j (E,J,\pi ,E^\mu_i,J^\mu_i,
\pi^\mu_i) T^\nu_o (E,J,\pi,E^\nu_m,J^\nu_m,\pi^\nu_m)} \over
{T_{tot} (E,J,\pi)}}
\label{hf}
\end{eqnarray}
for the reaction $i^\mu (j,o) m^\nu$ from the target
state $i^{\mu}$ to the exited state $m^{\nu}$ of the final nucleus, with
center of mass energy E$_{ij}$ and reduced mass $\mu _{ij}$. $J$ denotes the
spin, $E$ the excitation energy, and $\pi$ the parity of excited states.
When these properties are used  without subscripts they describe the compound
nucleus, subscripts refer to states of the participating nuclei in the
reaction $i^\mu (j,o) m^\nu$
and superscripts indicate the specific excited states. 
Experiments measure $\sum_{\nu} \sigma_{i} ^{0\nu} (j,o;E_{ij})$,
summed over all excited states of
the final nucleus, with the target in the ground state. Target states $\mu$ in
an astrophysical plasma are thermally populated and the astrophysical cross
section $\sigma^*_{i}(j,o)$ is given by
\begin{equation}
\label{astro}
\sigma^*_{i} (j,o;E_{ij}) = {\sum_\mu (2J^\mu_i+1) \exp(-E^\mu_i /kT)
\sum_\nu \sigma^{\mu \nu}_{i}(j,o;E_{ij}) \over \sum_\mu (2J^\mu_i+1)
 \exp(-E^\mu_i/kT)}\quad.
\end{equation}
The summation over $\nu$ replaces $T_o^{\nu}(E,J,\pi)$ in Eq.~\ref{hf} by
the total transmission coefficient
\begin{eqnarray}
T_o (E,J,\pi) & = &\sum^{\nu_m}_{\nu =0} 
T^\nu_o(E,J,\pi,E^\nu_m,J^\nu_m, \pi^\nu_m) \nonumber \\
& &+ \int\limits_{E^{\nu_m}_m}^{E-S_{m,o}} \sum_{J_m,\pi_m}
T_o(E,J,\pi,E_m,J_m,\pi_m)\rho(E_m,J_m,\pi_m) dE_m \quad.
\label{tot}
\end{eqnarray}
Here $S_{m,o}$ is the channel separation energy, and the summation over
excited 
states above the highest experimentally
known state $\nu_m$ is changed to an integration over the level density
$\rho$.
The summation over target states $\mu$ in Eq.~\ref{astro} has to be generalized
accordingly. 

In addition to the ingredients required for Eq.~\ref{hf}, like the
transmission coefficients for particles and photons, 
the width fluctuation corrections $W(j,o,J,\pi)$ have to be
employed. 
The important ingredients of statistical model calculations as indicated in
Eqs.~\ref{hf} through \ref{tot}
are the particle and $\gamma$-transmission coefficients $T$ and
the level density of excited states $\rho$. Therefore, the reliability of  
such calculations is determined by the accuracy with which these components 
can be evaluated. 

\section{Level Densities}

\subsection{The Back-Shifted Fermi-Gas Model}

\noindent
Considerable effort has been put into the improvement of the input for 
the statistical Hauser-Feshbach calculations (e.g.\ Cowan, Thielemann \& Truran
1991). However, 
the nuclear level density has given rise to the largest uncertainties in 
the description of nuclear reactions. For 
calculating the 
level densities in that context one does not only have to find reliable 
methods, but also computationally feasible ones. In dealing with 
thousands of nuclei one has to resort to simple models in order to 
minimize computer time.

Such a simple model
is the non-interacting Fermi-gas model introduced by Bethe (1936).
Mostly, the back-shifted Fermi-gas description, assuming an even 
distribution of odd and even parities, 
is used (Gilbert \& Cameron 1965):
\begin{equation}
\rho(U,J,\pi)={1 \over 2} f(U,J) \rho(U)\quad,
\end{equation}
with
\begin{eqnarray}
\rho(U)={1 \over \sqrt{2\pi}}{\sqrt{\pi} \over
12a^{1/4}}{\exp(2\sqrt{aU}) \over U^{5/4}}\ ,\qquad
f(U,J)={2J+1 \over 2\sigma^2} \exp\left({-J(J+1) \over
2\sigma^2}\right) \\
\sigma^2={\Theta_{\mathrm{rigid}} \over \hbar^2} \sqrt{U \over a}\ ,\qquad
\Theta_{\mathrm{rigid}}={2 \over 5}m_{\mathrm{u}}AR^2\ ,\qquad
U=E-\delta\quad. \nonumber
\end{eqnarray}
The spin dependence is determined by the spin cut-off parameter 
$\sigma$. Thus, the level density is dependent on only two parameters: 
the level density parameter $a$ and the backshift $\delta$, which 
determines the energy of the first excited state. 

Within this framework, the quality of level density predictions depends 
on the reliability of systematic estimates of $a$ and $\delta$. 
Gilbert \& Cameron (1965) were the first to identify an empirical
correlation of the level density parameter $a$ with experimental
shell corrections $S(N,Z)$ and of the backshift $\delta$ with experimental
pairing corrections.

An improved approach has to consider the energy dependence of the shell 
effects which are known to vanish at high excitation energies (Iljinov
et al.\ 1992), i.e.\ the thermal damping of shell effects.
Although, for astrophysical purposes only energies close to the particle 
separation thresholds have to be considered, an energy dependence can 
lead to a considerable improvement of the global fit. This is especially 
true for strongly bound nuclei close to magic numbers.

We use the following description (initially proposed 
by Ignatyuk, Smirenkin \& Tishin 1975; Ignatyuk, Istekov \& Smirenkin
1979)
for an energy-dependent level density parameter $a$:
\begin{equation}
\label{endepa}
a(U,Z,N)=\tilde{a}(A)\left[1+C(Z,N){f(U-\delta) \over 
U-\delta}\right]\quad,
\end{equation}
where
\begin{equation}
\tilde{a}(A)=\alpha A+\beta A^{2/3}
\end{equation}
and
\begin{equation}
f(U)=1-\exp(-\gamma U)\quad.
\end{equation}
The shape of the function $f(U)$ was found by approximation of
numerical microscopic calculations based on the shell 
model. Thus, we are left with three open parameters, 
namely $\alpha$, $\beta$, and $\gamma$. The values of these parameters 
can be determined by fitting experimental data.

Important input for these fits are the so-called shell correction 
$C(Z,N)$ and the pairing gap $\Delta$. These have to be taken 
from microscopic or macroscopic-microscopic nuclear mass formulae. 
Instead of assuming constant pairing or a fixed dependence on the mass
number $A$, we directly
determine the pairing gap $\Delta$ from mass differences of 
neighboring nuclei. Thus, for the neutron pairing gap $\Delta_{\mathrm{n}}$ one 
obtains (Wang et al.\ 1992)
\begin{equation}
\label{pair}
\Delta_{\mathrm{n}}(Z,N)={1 \over 4} \left[ 
E^G(Z,N-2)-3E^G(Z,N-1)+3E^G(Z,N)-E^G(Z,N+1)\right]\quad,
\end{equation}
where $E^G(Z,N)$ is the mass for the nucleus $(Z,N)$.
Similarly, the proton pairing gap $\Delta_{\mathrm{p}}$ can be 
calculated.

Previous attempts to find a global description of level densities used 
shell corrections derived from comparison of liquid-drop
masses with experiment ($C=S\equiv M_{\mathrm{exp}}-M_{\mathrm{LD}}$) or 
the ``empirical'' shell corrections $C(Z,N)=S(Z,N)$ given by Gilbert \&
Cameron (1965). 
A problem 
connected with the use of liquid-drop masses arises from the fact that 
there are different liquid-drop model parametrizations available in the
literature which produce quite different values for $S$.

We realized that one has to reconsider the meaning of the parameter $C$ 
in the level density formula (Eq.~\ref{endepa}). Actually, it should 
describe properties of a nucleus differing from the {\it spherical} macroscopic 
energy (i.e.\ mass), including properties which are vanishing at 
higher excitation 
energies. Therefore, the parameter $C$ should rather be identified with 
the so-called ``microscopic'' correction $E_{\mathrm{mic}}$. The mass of 
a nucleus with deformation $\epsilon$ can then be written as
\begin{equation}
M_{\mathrm{theo}}(\epsilon)=E_{\mathrm{mic}}(\epsilon)+E_{\mathrm{mac}}
(\mathrm{sphere})\quad.
\end{equation}
Alternatively, one can write 
\begin{equation}
M_{\mathrm{theo}}(\epsilon)=E_{\mathrm{mac}}(\epsilon)+E_{\mathrm{s+p}}
(\epsilon)\quad,
\end{equation}
with $E_{\mathrm{s+p}}$ being the shell-plus-pairing correction.
The confusion about the term ``microscopic correction'', which is 
sometimes ambiguously used in literature, is also pointed out 
in M\"oller et al.\ (1995). Thus, the above mentioned ambiguity of 
different liquid-drop 
parametrizations follows from the inclusion of partially deformation
dependent effects into the macroscopic part of the mass formula.

\subsection{Results}
\noindent
In our study we utilized the most recent mass formula by M\"oller et al.\
(1995), 
consistently also taking the microscopic corrections of that mass formula 
(calculated in a folded Yukawa shell model with Lipkin-Nogami pairing)
in order to determine the parameter $C(Z,N)$=$E_{\mathrm{mic}}$. The 
backshift $\delta$ was calculated by 
setting $\delta(Z,N)$=$\Delta_{\mathrm{n}}(Z,N)+\Delta_{\mathrm{p}}(Z,N)$
and using Eq.~\ref{pair}.
In order to obtain the parameters $\alpha$, $\beta$, and $\gamma$, we 
performed a fit to 
experimental data on s-wave neutron resonance spacing of 278 nuclei at 
the neutron separation energy. The 
data were taken from the compilation by Iljinov et al.\ (1992).

As a quantitative overall estimate of the agreement between calculation 
and experiment, one usually quotes the averaged 
ratio
\begin{equation}
g\equiv \left< {\rho_{\mathrm{calc}} \over \rho_{\mathrm{exp}}}\right> =
\exp \left[{1 \over n} \sum_{i=1}^{n}\left( \ln {\rho_{\mathrm{calc}}^i
\over \rho_{\mathrm{exp}}^i} \right)^2 \right]^{1/2}\quad,
\end{equation}
with $n$ being the number of nuclei for which level densities 
$\rho$ are experimentally known.

\begin{figure}
\psfig{file=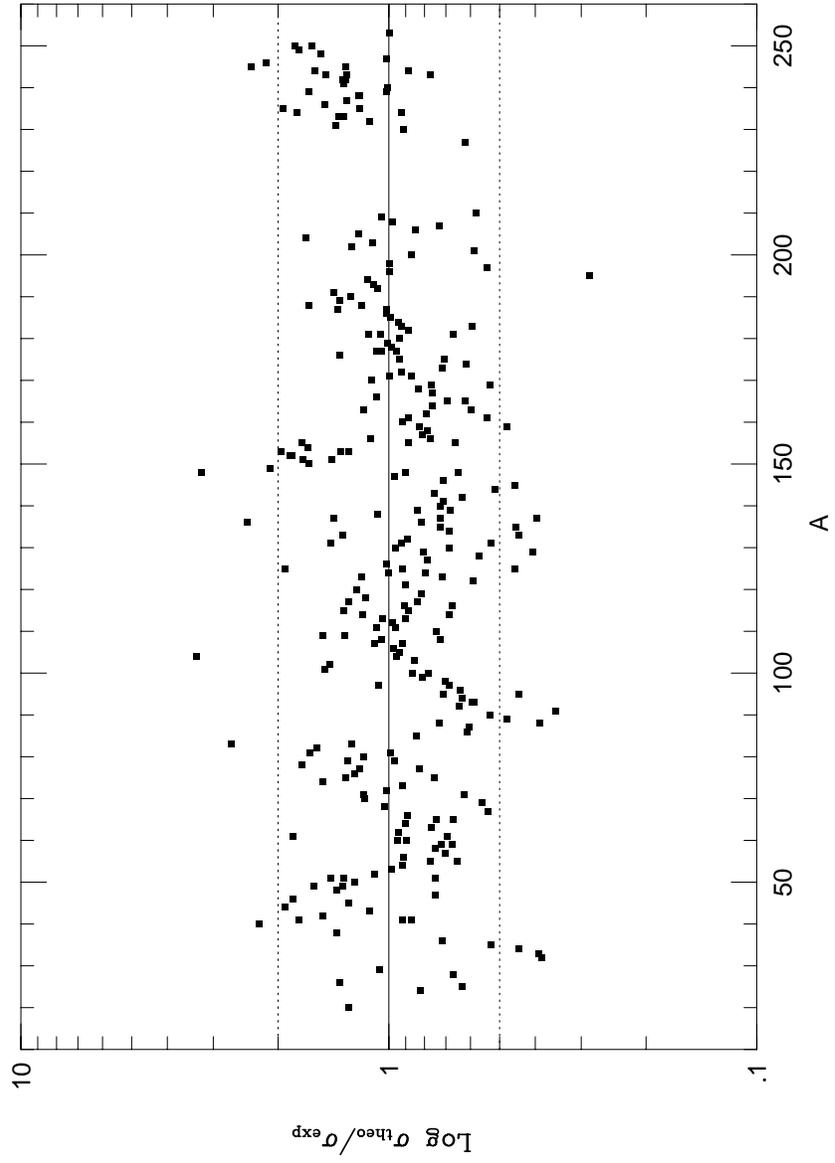,height=18cm,angle=180}
\caption{\label{figrat}Ratio of predicted to experimental (Iljinov et al. 1992)
level densities at the neutron separation energy.}
\end{figure}

As a best fit (when including known experimental masses where known),
we obtain an averaged ratio $g=1.5$ with the parameter 
values $\alpha=0.1336$, $\beta=-0.06712$, $\gamma=0.04862$. 
The ratios of 
experimental to predicted level densities for the nuclei considered is 
shown in Fig.~\ref{figrat}. As can be seen, for the majority of nuclei 
the absolute deviation is less than a factor of 2. This is a quite
satisfactory improvement over the theoretical level densities used previously.
Such a direct comparison was rarely shown in earlier work, mostly the level
density parameter $a$, entering exponentially into the level density, was
displayed.
Closely examining the nuclei with the 
largest deviations in our fit, we were not able to find any remaining
correlation of the deviation with separation energy (i.e. excitation
energy) or spin.
Different combinations of masses and microscopic corrections from other 
models (droplet model by Myers and Swiatecki, Cameron-Elkin mass formula 
and shell corrections, experimental values) were also tried but did not 
lead to better results.

As the parameters were fit to experimentally accessible nuclei close to 
the drip line, the question arises how reliable such a level density 
description can be for astrophysically interesting nuclei far from 
stability. Since the microscopic corrections (and the masses) are taken 
from a nuclear-structure model this leads to the question of the 
reliability of that model far off stability. There is, of course, no 
final conclusion regarding that point yet. However, recent investigations in 
astrophysics and nuclear physics have shown the robustness of the 
M\"oller et al.\ (1995) approach (Thielemann et al.\ 1994). 
Recently improved purely 
microscopic models have exhibited similar behavior towards the drip 
lines (Dobaczewski et al.\ 1994), but there are no large scale calculations 
over the 
whole chart of nuclei available yet. Therefore, the model used here 
is among the most 
reliable ones available at present.

\section{Applicability of the Statistical Model}

\noindent
With the help of a level density description one is also able to make 
statements on when and where the statistical model approach is valid. 
Generally speaking, in order to apply the model correctly a sufficiently 
large number of levels is needed in the relevant energy range which can
act as doorway states to form a compound nucleus. 
In the 
following we will discuss this with the aid of the above level density 
approach for neutron-, proton- and $\alpha$-induced reactions.
This section is intended to be a guide to a meaningful and correct 
application of the statistical model.

\subsection{Neutron-Induced Reactions}

\begin{figure}
\psfig{file=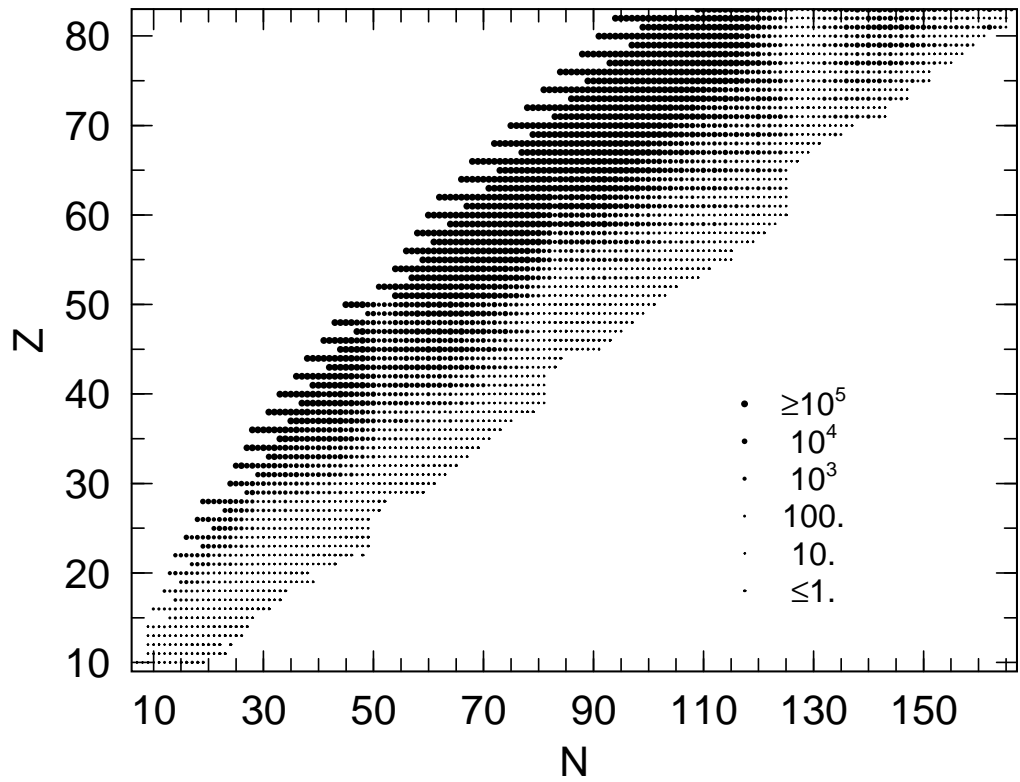,height=18cm}
\caption{\label{fig_sn}Level density (levels per MeV) at the neutron 
separation energy.}
\end{figure}

\noindent
In the case of neutron-induced reactions a criterion for the 
applicability can directly be derived from the level density. For 
astrophysical purposes the projectile energies are quite low by nuclear 
physics standards. Therefore, the relevant energies will lie very close 
to the neutron separation energy. Thus, one just has to consider the 
level density at this energy. As a rule-of-thumb it is usually said that 
there should at least be 10 levels per MeV for reliable statistical 
model calculations. The level densities at the appropriate neutron 
separation energies are shown in Fig.~\ref{fig_sn} (note that therefore 
the level density is plotted at a {\em different} energy for each 
nucleus). One can easily identify the magic neutron numbers by the drop 
in level density, as well as odd-even staggering effects. A general 
sharp drop is also found for nuclei close to the neutron drip line. For 
nuclei with such low level densities the statistical model cross 
sections will become very small and other processes might become 
important, such as direct reactions.

\subsection{Charged-Particle Induced Reactions}

\noindent
Contrary to
neutron-induced reactions, the knowledge of the level density at the
particle separation energy is not sufficient to say something about the
reliability of the statistical model because in this case also the
Coulomb barrier has to be taken into account. By folding the
Maxwell-Boltzmann velocity distribution of the projectiles at a given
temperature with the penetrability through the Coulomb barrier, the
so-called Gamow peak can be derived, in which most of the reactions will
take place. Location and width of the Gamow peak depend 
on the charges
of projectile and target, and on the temperature.

\begin{figure}
\psfig{file=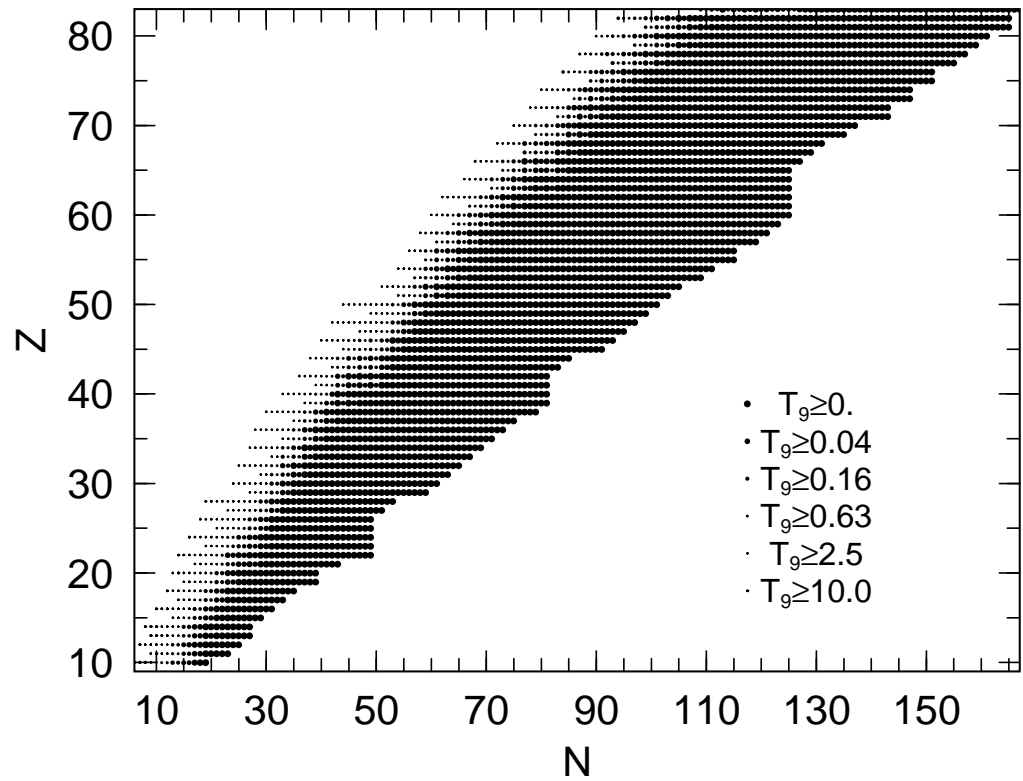,height=18cm}
\caption{\label{prot}Temperatures (in T$_9$) for which the statistical 
model can be used. Plotted is the compound nucleus of the proton-induced 
reaction p+Target (see text).}
\end{figure}

\begin{figure}
\psfig{file=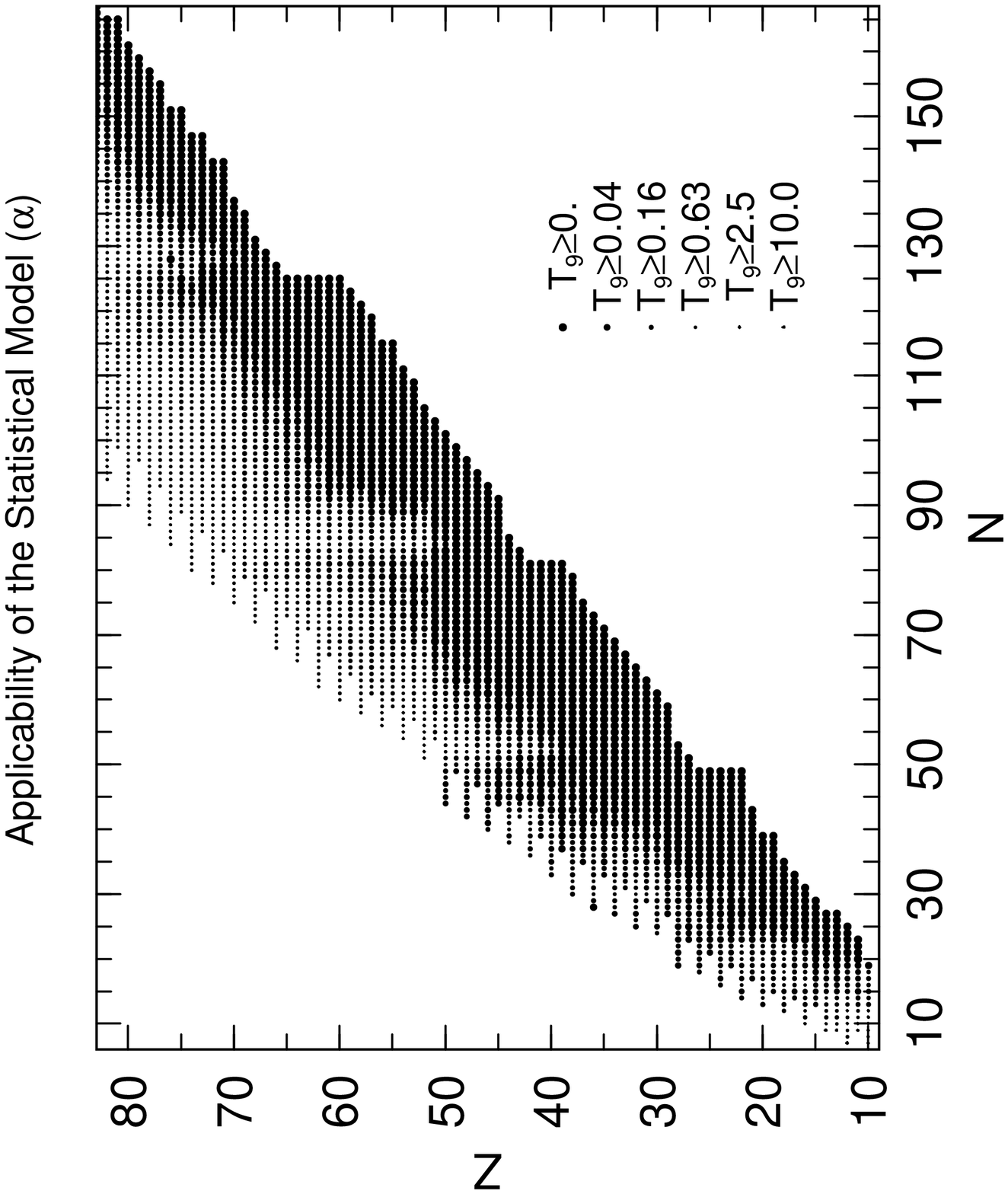,height=18cm}
\caption{\label{alph}Temperatures (in T$_9$) for which the statistical 
model can be
used. Plotted is the compound nucleus of the $\alpha$-induced reaction
$\alpha$+Target (see text).}
\end{figure}

Similar to the rule-of-thumb given above for neutron-induced reactions,
an estimate can be made of how many resonances have to be inside the
Gamow peak in order to still be able to numerically solve the integral
when calculating the reaction rate. Comparison with experimental cross sections
gives the result that one can assume a
value of about 3 resonances (levels). However, there are comparisons to
experimental data (Van Wormer et al.\ 1994) which have shown that in 
certain cases the experimental values could be reproduced with even fewer
levels. Nevertheless, to be on the safe side we assumed 3.5 levels
within the width (FWHM) of the Gamow peak as the necessary minimum in
the plots shown in Figs.~\ref{prot} and~\ref{alph}.

For proton-induced reactions Fig.~\ref{prot} should be used, whereas 
Fig.~\ref{alph} is for $\alpha$-induced reactions. These plots 
do not directly display level densities but rather show (again in a 
logarithmic scale)
the temperatures at which the statistical model can be used. That means
that for each nucleus (it is always the compound nucleus which is
plotted for the reaction) the temperature is given beyond which the 
above described condition
of 3.5 levels is fulfilled. This enables one to directly read from the
plot whether the statistical model cross section can be ``trusted'' or
whether single resonances or other processes (e.g.\ direct reactions) 
have also to be
considered. (However, this does not necessarily mean that the
statistical cross section is always negligible in the latter cases,
since the assumed condition is quite conservative).

\section{Summary}

\noindent
We were able to improve considerably the prediction of
nuclear level densities by employing an energy-dependent description for the
level density parameter $a$ and by correctly including microscopic
corrections.
All nuclei can now be
described with a single
parameter set consisting of just three parameters. 
The globally averaged deviation of prediction from experiment of about 
1.5 translates into somewhat smaller deviations of the final cross 
sections due to the dominance of transitions to states with low 
excitation energies.
This will also
make it worthwile to recalculate the cross sections and thermonuclear
rates for many astrophysically important reactions in the intermediate 
and heavy mass region.

We also presented a ``map'' as a guide for the application of 
the statistical model for neutron-, proton- and $\alpha$-induced 
reactions.
The above plots can give hints on when it is safe to use the statistical 
model approach and which nuclei have to be treated with special 
attention at a given temperature. Thus, information on which nuclei 
might be of special interest for an experimental investigation may also 
be extracted.
It should be noted that we used very conservative assumptions in deriving
the above criteria for the applicability of the statistical model.

\acknowledgements
This work was supported in part by the Swiss Nationalfonds.
TR is acknowledging support 
by an
APART fellowship from the Austrian Academy of Sciences.

\beginrefer

\refer Bethe, A.H.: 1936, \PR 50 332.

\refer Dobaczewski, J., Hamamoto, I., Nazarewicz, W., Sheikh, J.A.: 1994,
    {\it Phys.\ Rev.\ Lett.} {bf 72}, 981.

\refer Gilbert, A., Cameron, A.G.W.: 1965, {\it Can.\ J. Phys.} {\bf 43}, 1446.

\refer Hauser, W., Feshbach, H.: 1952, \PR A87 366.

\refer Holmes, J.A., Woosley, S.E., Fowler, W.A., Zimmerman, B.A.: 1976,
     {\it At. Data Nucl. Data Tables} {\bf 18}, 306.

\refer Ignatyuk, A.V., Smirenkin, G.N., Tishin, A.S.: 1975, {\it Yad.\ Phys.}
{\bf 21}, 485.

\refer Ignatyuk, A.V., Istekov, K.K., Smirenkin, G.N.: 1979, {\it Sov.\ J.
Nucl.\ Phys.} {\bf 29}, 450.

\refer Iljinov, A.S., et al.: 1992, {\it Nucl.\ Phys.} {\bf A543}, 517.

\refer Mahaux, C., Weidenm\"uller, H.A.: 1979, {\it Ann. Rev. Part. Nucl. Sci.}
{\bf 29}, 1.

\refer M\"oller, P., Nix, J.R., Myers, W.D., and Swiatecki, W.J.: 1995,
{\it At.\ Data Nucl.\ Data Tables} {\bf 59}, 185.


\refer Thielemann, F.-K., Arnould, M., Truran, J.W.: 1987,
    in {\it Advances in Nuclear Astrophysics}, eds. E. Vangioni-Flam
    et al., (Editions fronti\`eres: Gif sur Yvette), p.\ 525.

\refer Thielemann, F.-K., Kratz, K.-L., Pfeiffer, B., Rauscher, T.,
    van Wormer, L., Wiescher, M.: 1994,
    {Nucl.\ Phys.} {\bf A570}, 329c.

\refer Truran, J.W., Cameron, A.G.W., Gilbert, A.: 1966,
        {\it Can. J. Phys.} {\bf 44}, 563.

\refer Van Wormer, L., G\"orres, J., Iliadis, C., Wiescher, M., Thielemann,
F.-K.: 1994, \APJ 432 326.

\refer Wang, R.-P., Thielemann, F.-K., Feng, D.H., Wu, C.-L.: 1992,
    {\it Phys.\ Lett.} {\bf B284}, 196.

\refer Woosley, S.E., Fowler, W.A., Holmes, J.A., Zimmerman, B.A.:
1978, {\it At. Nucl. Data Tables} {\bf 22}, 371.

\endrefer
\end{document}